\documentclass[prl,showpacs,twocolumn,longbibliography]{revtex4-1}

\usepackage{xr}
\usepackage{hyperref}

\usepackage{color}
\usepackage[usenames,dvipsnames]{xcolor}
\usepackage{amsmath,amsthm,amssymb}
\usepackage{graphicx}
\usepackage{epsfig}
\usepackage{dcolumn}
\usepackage{bm}
\usepackage{mathrsfs}
\usepackage{multirow}
\usepackage[all]{xy}
\usepackage{pbox}
\usepackage{verbatim}

\DeclareMathOperator{\Tr}{Tr}

\newcommand{\er}[1]{Eq.~\eqref{#1}}
\newcommand{\ers}[2]{Eqs.~(\ref{#1}-\ref{#2})}

\def\(({\left(}
\def\)){\right)}
\def\[[{\left[}
\def\]]{\right]}

\newcommand*{\QEDB}{\hfill\ensuremath{\square}}%

\newcommand{\be}{\begin{equation}}
\newcommand{\ee}{\end{equation}}
\newcommand{\ben}{\begin{eqnarray}}
\newcommand{\een}{\end{eqnarray}}
\newcommand{\beq}{\begin{equation}}
\newcommand{\eeq}{\end{equation}}

\newcommand{\la}{\langle}
\newcommand{\ra}{\rangle}

\newcommand{\e}{{\text{e}}}

\begin{document}

\title{Making rare events typical in Markovian open quantum systems}

\author{Federico Carollo, Juan P. Garrahan, Igor Lesanovsky and Carlos P\'erez-Espigares}
\affiliation{School of Physics and Astronomy}
\affiliation{Centre for the Mathematics and Theoretical Physics of Quantum Non-Equilibrium Systems,
University of Nottingham, Nottingham, NG7 2RD, UK}

\date{\today}

\begin{abstract}
Large dynamical fluctuations - atypical realizations of the dynamics sustained over long periods of time - can play a fundamental role in determining the properties of collective behavior of both classical and quantum non-equilibrium systems. 
Rare dynamical fluctuations, however, occur with a probability that often decays exponentially in their time extent, thus making them difficult to be directly observed and exploited in experiments. 
Here, using methods from dynamical large deviations, we explain how rare dynamics of a given (Markovian) open quantum system can always be obtained from the typical realizations of an alternative (also Markovian) system.  The correspondence between these two sets of realizations can be used to engineer and control open quantum systems with a desired statistics ``on demand''. We illustrate these ideas by studying the photon-emission behaviour of a three-qubit system which displays a sharp dynamical crossover between active and inactive dynamical phases.
\end{abstract}

\maketitle 
\noindent {\bf \em Introduction.} 
The exploration and control of quantum matter far from equilibrium is a current theme in physics. This interest is rooted in progress in realizing and probing many-body dynamics with ensembles of cold atoms or trapped ions; for a few recent examples see \cite{Barreiro2011,Schauss2012,Schachenmayer2013,Eisert2015,Islam2015,Schreiber2015,Valado2016,Bohnet2016,Zhang2017,Zhang2017b,lienhard2018a,bernien2017,guardado2018}. Beyond enabling the study of ground states and unitary evolution, these systems also allow a controlled coupling to an environment. Collective dynamics, e.g.\ due to a competition between interactions and external driving, may then be monitored through the quanta emitted from the system into the environment \cite{Plenio1998,Breuer2002,Gardiner2004,Daley2014}. 

\begin{figure}[t]
\centering
\includegraphics[scale=0.37]{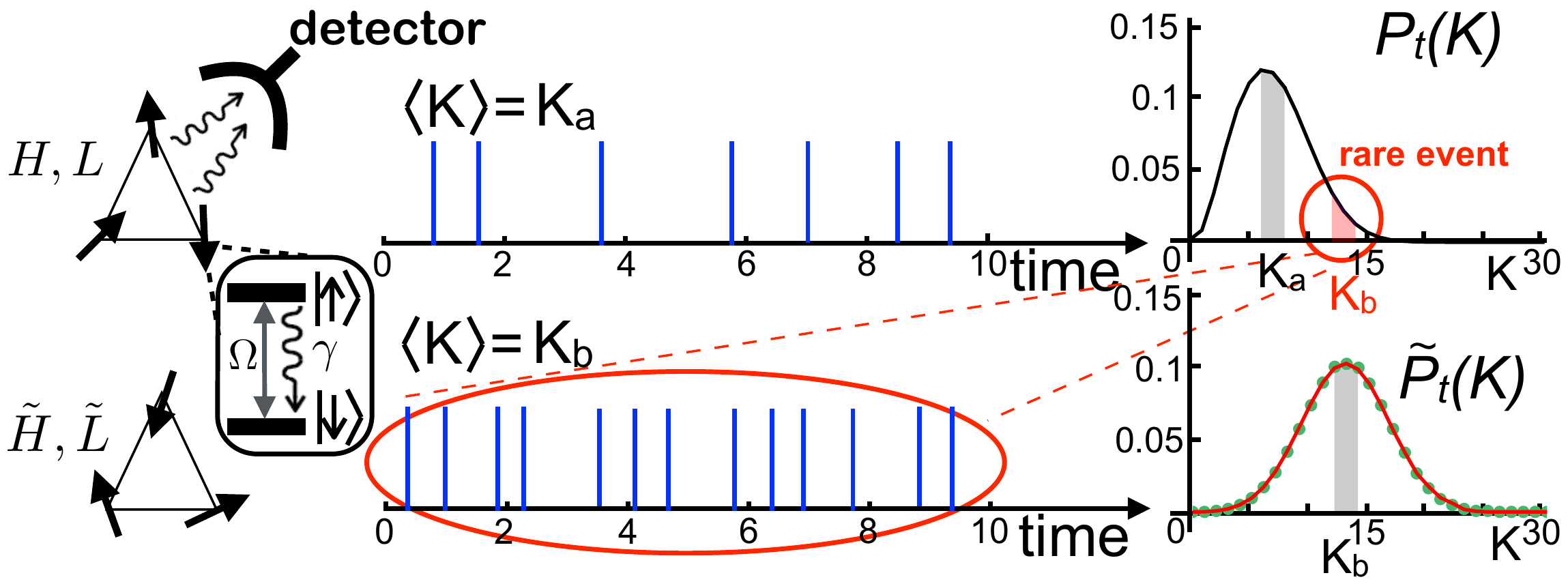}
\caption{\small Three-qubit system with dynamics described by Eqs. \eqref{Ham}-\eqref{jump} with $\Omega=V=\gamma=1$, and initial state with all qubits in the excited state. A photon is emitted whenever the excited state $|\!\!\uparrow \ra$ decays into the ground state $|\!\!\downarrow \ra$. Photons are detected (blue vertical lines) and their statistics $P_t(K)$, with $K$ being the total amount of emitted photons, is sampled up to time $t=10$. {\bf Top:} Typical records of emissions show that the average number of emitted photons is $K_a\sim 7$. {\bf Bottom:} Modified quantum system for which the rare event $K_b=13$ of the original dynamics becomes typical, as shown by $\tilde{P}_t(K)$: green points are obtained by sampling quantum trajectories of the modified dynamics $\tilde H, \tilde L$ discussed in the last section, while the red solid line corresponds to the biased ensemble of probabilities \eqref{pisa} for $s=-0.5$.}
\label{Fig1v1}
\end{figure}

Often, such emission output is characterized by a mostly unstructured sequence of events (e.g.\  photons emitted from atoms), but with occasional rare periods of ordered sequences or bursts of photons. For instance, consider a three-qubit system coupled to an environment and driven from an external laser field, as shown Fig.~\ref{Fig1v1}: while the typical number of emitted photons is $K_a$, a rare dynamical fluctuation is characterized by the emission of e.g. $K_b\sim2\, K_a$ photons. Unfortunately, experimental observation and control of these atypical events is in general out of reach as their occurrence probability is suppressed exponentially in their time duration. Thus, the question that naturally arises is whether it is possible to modify the quantum system in such a way that these interesting fluctuations are becoming the typical dynamics, as sketched in Fig.~\ref{Fig1v1}.

Here we explain how to engineer a quantum system whose typical dynamics is the same as that corresponding to rare dynamical fluctuations of a given original system. 
The underlying theoretical framework to tackle this problem is provided by large deviation (LD) theory \cite{Eckmann1985,den2008large,Touchette2009} as applied to open quantum systems \cite{Garrahan2010}. We show here how the LD formalism allows - if the original dynamics is described by a Lindbladian master equation \cite{Gorini1976,Lindblad1976,Gardiner2004}- to obtain the Hamiltonian and the jump operators of the new quantum system (sometimes, in the classical context, called ``auxiliary'' \cite{jack2010large} or ``driven'' \cite{Chetrite2015}) whose statistics of events corresponds to a precise rare behavior of the original system. This construction can be seen as the quantum counterpart \cite{Garrahan2010} of the so-called generalized Doob transform \cite{jack2010large,doob2012classical,belitsky2013,harris2013,Chetrite2013,Nemoto2014,Chetrite2015,Chetrite2015variational,Garrahan2016}. While we focus mostly on the mapping at long times, we also provide the quantum generalisation of the time-dependent Doob transform which makes the correspondence of the statistics valid for all times \cite{Chetrite2015,Garrahan2016}. This means that biasing can be even achieved for \emph{transient dynamics}, which may be more easily accessible experimentally than stationary behaviors. 

We illustrate our ideas with a three-qubit system characterized by a pronounced crossover at the level of fluctuations between dynamical regimes of atypically low and atypically high photon emission rates. For these rare inactive and active behaviors of the original system, we derive a new dynamics where they become typical. In the inactive case, the new system has collective jump operators that make photon emission dynamically constrained. In contrast, the active case corresponds to dynamics where qubits are almost independent. 

Our findings show the possibility of tuning many-body quantum systems to dynamical critical points or phase transitions \cite{Carr2013,Ivanov2015,Weller2016,Letscher2017,Fernandez2017,wade2017,Raghunandan2018}.
In addition, the proposed engineering of open quantum dynamics may become of practical interest for realizing quantum devices relying on streams of photons, electrons or ions \cite{Keller2004,Lounis2005,Barros2009,Schnitzler2009,Sahin2017} with not only a controlled average output but also with tailored fluctuations.

\smallskip

\noindent {\bf \em Open quantum dynamics and thermodynamics of quantum trajectories.} 
The LD formalism provides a method to study ensembles of dynamical trajectories using an approach equivalent to that of standard statistical mechanics for ensembles of configurations in equilibrium. The state $\rho_t$ of the Markovian quantum system we consider here evolves according to the master equation $\dot{\rho}_t=\mathcal{L}[\rho_t]$, where the Lindbladian generator is given by \cite{Gorini1976,Lindblad1976,Gardiner2004}
\begin{equation}
\mathcal{L}[~\!\!\cdot~\!\!]=-i[H,\cdot]+\sum_{\mu=1}^{N_J}\left(L_\mu\cdot L_\mu^\dagger-\frac{1}{2}\left\{L_\mu^\dagger L_\mu,\cdot\right\}\right)\, .
\label{L}
\end{equation}
Here, $H=H^\dagger$ is the system Hamiltonian, $L_\mu$ ($\mu=1,...,N_J$) are the jump operators and $\{\cdot,\cdot\}$ stands for the anticommutator. Each of these jump operators corresponds to a specific type of event detected in the environment. For example in Fig.~\ref{Fig1v1} this describes the emission of a photon, but in general can be e.g.~bath quanta emission/absorption, or particle injection/ejection.  

Assuming ideal detectors for these events, one is able to obtain their time records or \emph{quantum trajectories} \cite{Garrahan2010}. To any trajectory one can then associate a vector of outcomes $\vec{K}$ whose components are the total counts of the detected events of each kind (or linear combinations of these total counts). 
The corresponding probability distribution $P_t(\vec{K})$, in the quantum framework, is given by $P_t(\vec{K})=\Tr(\rho_t^{\vec{K}})$, where $\rho_t^{\vec{K}}$ is the state of the system obtained by {\em conditioning} trajectories on having exactly outcomes ${\vec{K}}$ up to time $t$ \cite{Zoller1987,Garrahan2010,Gammelmark2013}. In the long-time limit, these probabilities are assumed to satisfy a LD principle \cite{Touchette2009}, $P_t(\vec{K})\approx \e^{-t\, \phi(\vec{k})}$, with $\vec{k}=\vec{K}/t$. The function $\phi(\vec{k})$ is known as the LD {\em function} (LDF), and is minimized at $\langle {\vec{k}} \rangle$. This asymptotic form shows that dynamical realizations sustaining atypical outcomes away from $\langle {\vec{k}} \rangle$ are exponentially suppressed in time.

Formally, the standard way of {\em tilting} or biasing these probabilities towards desired values of the outcomes is by defining an ensemble of trajectories such that \cite{Garrahan2010,Garrahan2007b,Chetrite2013,Chetrite2015}
\begin{equation}
P_t^{\vec{s}}(\vec{K})=\frac{\e^{-\vec{s}\cdot \vec{K}}}{\mathcal{Z}_t(\vec{s})}\, P_t(\vec{K}),
\label{pisa}
\end{equation}
\begin{equation}
\mathcal{Z}_t(\vec{s}):=\sum_{\vec{K}}\e^{-\vec{s}\cdot\vec{K}}\, P_t(\vec{K})\, ,
\label{pisb}
\end{equation}
where the entries of $\vec{s}$ are parameters conjugated to the observables which quantify the strength of the bias. For example in Fig.~\ref{Fig1v1} an appropriate biasing transforms the probability $P_t(K)$ with average emission $K_a$, into the probability ${\tilde P}_t(K)$ with average $K_b$. The same information is contained in the moment generating function \eqref{pisb}. The change from the conditioned ensemble where all trajectories have fixed $\vec{K}$ to the biased one defined by \er{pisa}, where only $\langle \vec{K} \rangle_{\vec{s}}$ is fixed, is analogous to the change from the microcanonical to the canonical ensemble in equilibrium statistical mechanics \cite{Chetrite2013,Touchette2015}. 
Trajectories with outcomes $\vec{K}$ different from the typical ones can be favored or suppressed, varying the vector of the conjugate parameters $\vec{s}$.  At long times, the whole statistics of the time-averaged observables $\vec{k}$ in these biased ensembles is encoded in the scaled cumulant generating function (SCGF) $\theta(\vec{s})= \lim_{t \to \infty} t^{-1}\log\mathcal{Z}_t(\vec{s})$ \cite{Garrahan2007b}. 

The parameters $\vec{s}$, however, are not physical quantities that can be directly tuned in experiments \cite{Brandner2017}. In order to practically access the rare dynamical behavior that mathematically is controlled by $\vec{s} \neq 0$, we show below how to define an alternative system in terms of a different Hamiltonian and jump operators whose unbiased (and therefore physical) dynamics has biased probabilities given by \er{pisa}.

\smallskip

\noindent  {\bf \em Example: three-qubit system.} To make our ideas concrete, we apply them to the open three-qubit system depicted in Fig.~\ref{Fig1v1}, which can be experimentally realized, for example, by means of trapped ions \cite{Kim2010,Bohnet2016} or Rydberg atoms \cite{Schauss2012,Labuhn2016,bernien2017}. The quantum dynamics is described by a Lindblad generator of the form \eqref{L}, with Hamiltonian
\begin{equation}
H=\Omega\sum_{k=1}^3\sigma_{\rm x}^{(k)}+V\hspace{-0.2cm}\sum_{k>h=1}^3\sigma_{\rm z}^{(k)}\sigma_{\rm z}^{(h)},
\label{Ham}
\end{equation}
and jump operators describing independent decay (and photon emission) from each qubit ($k=1,2,3$) at rate $\gamma$
\begin{equation}
L_k = \sqrt{\gamma} \sigma_{-}^{(k)} \, .
\label{jump}
\end{equation}
Here $\sigma_{\alpha}^{(k)}$ ($\alpha = 0,x,y,z$) are the Pauli matrices for $k$-th qubit with $\sigma_0$ the identity and $\sigma_{\pm}=\frac{1}{2}\left(\sigma_{\rm x}\pm i\sigma_{\rm y}\right)$.

\begin{figure*}
\centering
\includegraphics[scale=0.525]{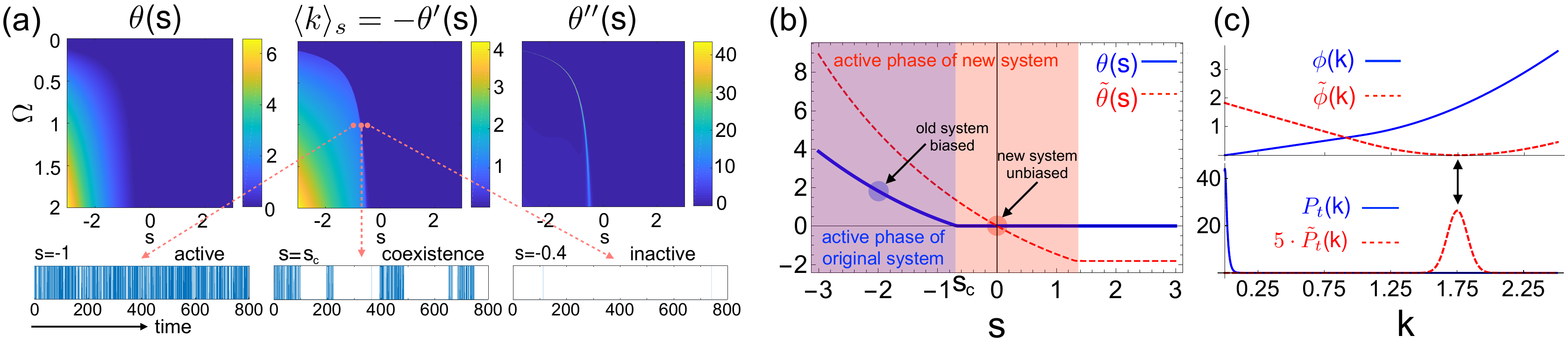}
\caption{\small {\bf (a) Top:} Scaled cumulant generating function (SCGF) of the original three-qubit system and its first two derivatives for $V=10$. The first order derivative displays an abrupt cross-over from an inactive phase to an active one. {\bf Bottom:} Emission time record for $\Omega=1$: each blue line corresponds to the detection of a photon. Trajectories are obtained performing quantum jump Monte Carlo for the modified dynamics. {\bf (b)} SCGF of the original system (solid blue line) and the modified system (red dashed line) for $s_0=-2$ and $\Omega=1$. {\bf (c) Top:} Original LDF (solid blue line) with average activity $\la k \ra_0=2\times 10^{-3}$ and the modified one (red dashed line) with average activity $\la k \ra_{s_0}=1.75$. {\bf Bottom:} Corresponding probability distributions for $t=100$, whose maximum is at the average activity in each case.
}
\label{Ftheta}
\end{figure*} 

In Fig.~\ref{Ftheta}(a) we show the SCGF $\theta(s)$, together with the mean emission rate $\la k\ra_s=-\theta'(s)$ and the fluctuations $\theta''(s)$ as functions of $\Omega$ and $s$. For almost all choices of $\Omega$ we observe that the mean emission rate $\la k\ra_s=-\theta'(s)$ displays a sharp change from an active to an inactive regime, in which trajectories are characterised by either a large or small photon count rate. Directly at the crossover, $s=s_c\approx -0.68$, trajectories are a mixture of both active and inactive ones \cite{Garrahan2010}. Here the two dynamical phases coexist \cite{Hedges2009,Ates2012}. In order to access the active phase as well as the coexistence region, the system needs to be biased. This means that the Hamiltonian and the jump operators are adjusted such that the rare dynamics that is currently assumed at values of $s$ away from $0$ is becoming the typical dynamics of the  new system. The technical details of this procedure are discussed in the following section, but we already illustrate the results in Figs.~\ref{Ftheta}(b,c). Here we bias the three-qubit system such that the dynamics is in the active phase. In Fig.~\ref{Ftheta}(b) we see that the new SCGF,  $\tilde{\theta}(s)$, is the original SCGF but shifted in a way that the crossover (formerly at $s_c<0$) is now found at $s>0$. Thus, the new system is indeed within the active phase which is also reflected in the distribution function of the photon counts, Fig.~\ref{Ftheta}(c).


\smallskip

\noindent {\bf \em Quantum Doob Transform.} We will now show how the desired transformation of the Hamiltonian and jump operators is achieved. The starting point is the tilted generator \cite{Garrahan2010,Znidaric2014,Manzano2017,Monthus2017b,Carollo2017}
\begin{equation}
\mathcal{L}_{\vec{s}_0}[\cdot]=-i[H,(\cdot)]+\sum_{\mu=1}^{N_J}\left(\e^{f_\mu(\vec{s}_0)}\, L_\mu(\cdot)L_\mu^\dagger-\frac{1}{2}\left\{L_\mu^\dagger L_\mu,(\cdot)\right\}\right)\, ,
\label{TiltL}
\end{equation}
where the functions $\{f_\mu(\vec{s}_0)\}_{\mu=1}^{N_J}$, such that $\forall \mu$, $f_\mu(\vec{s}_0=0)=0$, are linear in $\vec{s}_0$ and encode the dependence of the observables of interest on the various kind of jumps. The generator (\ref{TiltL}) generates the ensemble associated with the probabilities $P_t^{\vec{s}_0}(\vec{K})$, cf.\ \er{pisa}, but it does not represent a physical dynamics, i.e. it is not trace-preserving at $\vec{s}_0 \neq 0$. 

At long times, the link to the  probabilities $P_t^{\vec{s}_0}(\vec{K})$ is through the largest eigenvalue of 
$\mathcal{L}_{\vec{s}_0}$ which gives the SCGF $\theta(\vec{s}_0)$ associated to the biased ensemble. To this leading eigenvalue belong left and right eigenmatrices, 
$\mathcal{L}^{*}_{\vec{s}_0}[\ell_{\vec{s}_0}]=\theta(\vec{s}_0)\, \ell_{\vec{s}_0}$
and
$\mathcal{L}_{\vec{s}_0}[r_{\vec{s}_0}]=\theta(\vec{s}_0)\, r_{\vec{s}_0}$, 
normalized such that $\Tr\left(\ell_{\vec{s}_0} \, r_{\vec{s}_0}\right)=\Tr\left(r_{\vec{s}_0}\right)=1$, with $\mathcal{L}_{\vec{s}_0}^*$ the dual map acting on the system operators.

The remaining task to convert \er{TiltL} to a proper Lindbladian generator, i.e. to remedy the lack of trace-preservation. To this end we define, generalizing Ref.\ \cite{Garrahan2010}, the operator
$$
\tilde{\mathcal{W}}_{\vec{s},\vec{s}_0}[\cdot]=\ell_{\vec{s}_0}^{1/2}\mathcal{L}_{\vec{s}+\vec{s}_0}\left[\ell_{\vec{s}_0}^{-1/2}\, ~(\cdot)\, \ell_{\vec{s}_0}^{-1/2}\right]\ell_{\vec{s}_0}^{1/2}-\theta(\vec{s}_0)~(\cdot) \, .
$$
This is a completely positive map obeying $\tilde{\mathcal{W}}_{\vec{s}=0,\vec{s}_0}^*[{\bf 1}]=0$. We can rewrite this in a more convenient way \cite{SM}:
\begin{align}
\tilde{\mathcal{W}}_{\vec{s},\vec{s}_0}[\cdot] &= -i[{\tilde H}_{\vec{s}_0},(\cdot)]+
\label{TiltDoob} 
\\
& +\sum_{\mu=1}^{N_J}\left(\e^{f_\mu(\vec{s})}\, {\tilde L}_\mu^{\vec{s}_0}(\cdot) {\tilde L}_\mu^{\vec{s}_0\dagger}-\frac{1}{2}\left\{{\tilde L}_\mu^{\vec{s}_0\dagger} {\tilde L}_\mu^{\vec{s}_0},(\cdot)\right\}\right)
\nonumber
\end{align}
with the modified jump operators
\begin{align}
{\tilde L}_\mu^{\vec{s}_0} &= \e^{\frac{1}{2}f_\mu(\vec{s}_0)}\ell_{\vec{s}_0}^{1/2}\, L_\mu\, \ell_{\vec{s}_0}^{-1/2}
\label{LDoob}
\end{align}
and the modified Hamiltonian
\begin{align}
{\tilde H}_{\vec{s}_0} &= \frac{1}{2}\ell_{\vec{s}_0}^{1/2}\left(H-\frac{i}{2}\sum_{\mu=1}^{N_J}L_\mu^\dagger L_\mu\right)\ell_{\vec{s}_0}^{-1/2}+\text{h.c.}
\label{HDoob}
\end{align}
For $\vec{s}=0$, the map is clearly in Lindblad form, and its steady-state is given by $\rho^{\vec{s}_0}_\infty=\ell^{1/2}_{\vec{s}_0}r_{\vec{s}_0}\ell^{1/2}_{\vec{s}_0}$. 

We shall prove now that the map $\tilde{\mathcal{W}}_{\vec{s}=0,\vec{s}_0}$ generates trajectories whose statistics, for long times, is given by $P_t^{\vec{s}_0}(\vec{K})\approx e^{-t(\phi(\vec{k})+\theta(\vec{s}_0)+\vec{k}\cdot\vec{s}_0)}$, which is nothing but the asymptotic form of Eq.~\eqref{pisa}.


\noindent {\bf Theorem.} \emph{The dynamical generator $\tilde{\mathcal{W}}_{\vec{s}=0,\vec{s}_0}$ describes a quantum system, for which the long-time statistics of the time-averaged observables ${\vec k}={\vec K}/t$ is characterized by the following SCGF and LDF:} 
\begin{equation}
\tilde{\theta}(\vec{s})=\theta(\vec{s}+\vec{s}_0)-\theta(\vec{s}_0),\,\, \,\,
\tilde{\phi}(\vec{k})=\phi(\vec{k})+\theta(\vec{s}_0)+\vec{k}\cdot\vec{s}_0\, .
\label{C1}
\end{equation}
\noindent {\em Proof:} $\tilde{\mathcal{W}}_{\vec{s}=0,\vec{s}_0}$ is a Lindblad map, and thus represents a quantum generator. Moreover, with respect to the observables ${\vec K}$, $\tilde{\mathcal{W}}_{\vec{s},\vec{s}_0}$ is the tilted-operator of  $\tilde{\mathcal{W}}_{\vec{s}=0,\vec{s}_0}$. Recalling that $r_{\vec{s}}$ is the right eigenmatrix corresponding to the largest eigenvalue of the map $\mathcal{L}_{\vec{s}}$, the eigenvalue with the largest real part of $\tilde{\mathcal{W}}_{\vec{s},\vec{s}_0}$ can be simply found noticing that
$
\tilde{\mathcal{W}}_{\vec{s},\vec{s}_0}[R]=\tilde{\theta}(\vec{s})\, R\, ,
$
with $R=\ell_{\vec{s}_{0}}^{1/2}r_{\vec{s}+\vec{s}_0}\ell_{\vec{s}_{0}}^{1/2}$.
This map is not uniquely defined: the same result is obtained for the generator $\mathcal{U}\circ\tilde{\mathcal{W}}_{\vec{s},\vec{s}_0}\circ\mathcal{U}^{-1}$, with eigenmatrix $\mathcal{U}[R]$, where $\mathcal{U}$ is a unitary map $\mathcal{U}[\cdot]=U\cdot U^\dagger$.
The LDF $\tilde{\phi}(\vec{k})$ is related to $\tilde{\theta}(\vec{s})$, via the Legendre transform 
$
\tilde{\phi}(\vec{k})=\max_{\vec{s}}[-\vec{k}\cdot\vec{s}-\tilde{\theta}(\vec{s})]
$.
Adding and subtracting $\vec{k}\cdot\vec{s}_0$, the above relation can be rewritten as 
$$
\tilde{\phi}(\vec{k})=\max_{\vec{s}}[-\vec{k}\cdot(\vec{s}+\vec{s}_0)-\theta(\vec{s}+\vec{s}_0)]+\theta(\vec{s}_0)+\vec{k}\cdot\vec{s}_0\, .
$$
Changing variables in the maximization procedure, and noticing that $\phi(\vec{k}):=\max_{\vec{s}}[-\vec{k}\cdot\vec{s}-\theta(\vec{s})]$, one obtains the second relation in \eqref{C1}.
\QEDB

The SCGF of the modified dynamics is just a shift of the original one, as depicted in Fig.~\ref{Ftheta}(b). We stress that this new SCGF is associated with the  LDF $\tilde{\phi}(\vec{k})$ which clearly shows 
that the biased probability $P_t^{\vec{s}_0}(\vec{K})$ is now given by the typical behavior of the physical dynamics $\tilde{\mathcal{W}}_{\vec{s}=0,\vec{s}_0}$ as displayed in Fig.~\ref{Ftheta}(c). The Doob transformation enables thus the tuning of the system to any point in the dynamical phase diagram.

\smallskip

\begin{figure}[t]
\centering
\includegraphics[scale=0.24]{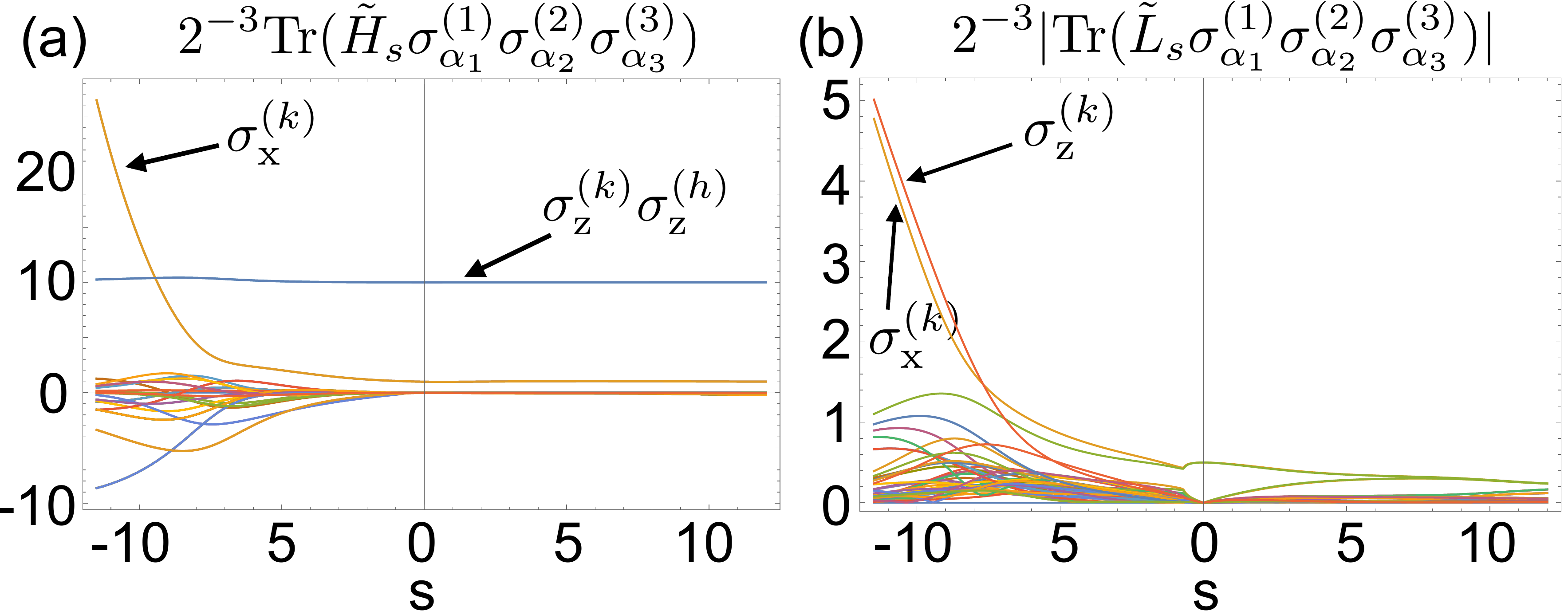}\caption{\small Coefficients of the operator decomposition of the new Hamiltonian and jump operators on the Pauli matrix basis as a function of $s$. {\bf (a)}  Real coefficients of ${\tilde H}_{s}$. {\bf (b)} Absolute value of the complex coefficients of ${\tilde L}_k^s$. For $s_0=-11.5$ (active phase), $\Omega'\sim 26.5$ and $\gamma'\sim4.8$ give an emission rate of $\la k\ra\sim 55.4$. For $s_0=12$ (inactive phase), with $\alpha=0.95$, $\beta=0.4$ the typical emission rate is reduced to $\la k\ra\sim 10^{-5}$. }
\label{Doob}
\end{figure}

\noindent  {\bf \em Engineering an active and inactive dynamics.}
Let us now return to the three-qubit system. We have previously seen that we can indeed construct a new dynamics that brings the system from the inactive into the active phase. The question we are asking now is what new terms in the Hamiltonian and Lindblad operators are generated by the Doob transform that make the system undergo this change in dynamics. To answer this question we decompose the Hamiltonian and jump operators ${\tilde H}_{s_0},{\tilde L}_{\mu}^{s_0}$ into the Pauli matrix basis. In this way it is possible to identify relevant terms of the Doob dynamics. In Fig.~\ref{Doob}(a), it is clear that, for large negative $s$, the main contribution to the Hamiltonian is given by $\sigma_{\rm x}^{(k)}$. Analogously [see Fig.~\ref{Doob}(b)], when biasing deep into the active phase, jump operators are dominated by the terms $\sigma_{\rm x}^{(k)}$ and $\sigma_{\rm z}^{(k)}$, and one can approximate them as ${\tilde L}_{k}^{s_0}\approx \sqrt{\gamma'}(\sigma_{\rm x}^{(k)}-i\sigma_{\rm z}^{(k)})$.
Unitarily rotating the basis in such a way that $x\to x$, $y\to-z$, and $z\to y$, the biased dynamics is approximately implemented by jump operators $\{2\sqrt{\gamma'}\, \sigma_-^{(k)}\}_k$, and Hamiltonian 
$
\tilde{H}_{s_0}\approx\Omega' \sum_{k}\sigma_{\rm x}^{(k)},
$
where $\Omega'$ and $\gamma'$ depend on the value of $s_0$.
This shows that in order to produce high emission rates the system becomes non-interacting. 

We now look at the (inactive) dynamics biased to large positive $s$.
The Hamiltonian remains virtually unchanged compared to the original one [see Fig.~\ref{Doob}(a)], whereas Fig.~\ref{Doob}(b) does not allow to identify a clear dominant contribution to the jump operators. However, inspecting directly the matrix elements, we find that they are well approximated by ${\tilde L}_k^{s_0}\approx \alpha \prod_h \sigma_-^{(h)}+\beta\sigma_-^{(k)}\prod_{h\neq k}(1-\sigma_{\rm z}^{(h)})$. This shows that activity is not simply reduced by reducing the decay rate $\gamma$, but rather by constraining the photon emission: from ${\tilde L}_k^{s_0}$ we see that photons can be emitted either if all qubits are in the excited state, or if one is in the excited state and the others in the ground state. Note, that such collective jump operators may be challenging to implement experimentally. They may be realized as an emergent dynamics of a strongly interacting system in a perturbative limit --- similar to kinetic constraints in dissipative Rydberg gases \cite{lesanovsky2013,gutierrez2017,letscher2017a}. A further possibility is to engineer them using digital quantum simulation protocols which have been demonstrated in ion traps \cite{lanyon2011}.


\smallskip

\noindent  {\bf \em Time-dependent quantum Doob transform.} So far we have focused on the stationary state dynamics, but now we ask whether it is in principle possible to generate the biased dynamics also during the approach to stationarity. This is achieved by a time-dependent Doob transform (see Refs. \cite{Chetrite2015,Garrahan2016} for classical stochastic systems), for which the statistics of events is given by Eq.~\eqref{pisa} {\em for all observation times} $t_{\rm f}$. 
The tilted evolution up to the final time $t_{\rm f}$ can be divided into $N$ time steps, $e^{t_{\rm f}\mathcal{L}_{\vec{s}_0}}=\prod_{k=1}^N e^{\delta t\mathcal{L}_{\vec{s}_0}}$, with $\delta t=t_{\rm f}/N$. Introducing  the map $g_t[X]=G_tXG_t$ and its inverse $g_t^{-1}$, being $G_t$ an arbitrary Hermitian time-dependent operator, we rewrite the tilted evolution as 
$
e^{t_{\rm f}\mathcal{L}_{\vec{s}_0}}=g_{t_{\rm f}}^{-1}\circ \left(\prod_{k=1}^N g_{t_k}\circ e^{\delta t\mathcal{L}_{\vec{s}_0}}\circ g_{t_k-\delta t}^{-1}\right)\circ g_{0}
$.
The product is from largest to smallest times $t_k=(N+1-k)\delta t$. For $\delta t\ll1$, one has  
$g_{t_k}\circ e^{\delta t\mathcal{L}_{\vec{s}_0}}\circ g_{t_k-\delta t}^{-1}\approx e^{\int_{t_k-\delta t}^{t_k}du\, \tilde{\mathcal{L}}_u}$, with
$$
\tilde{\mathcal{L}}_t[\rho]=G_t\mathcal{L}_{\vec{s}_0}\left[G_t^{-1}\rho\,  G_t^{-1}\right]G_t +\dot{G}_tG_t^{-1}\rho +\rho \, G_t^{-1}\dot{G}_t\, .
$$

\begin{figure*}[t!]
\includegraphics[scale=0.73]{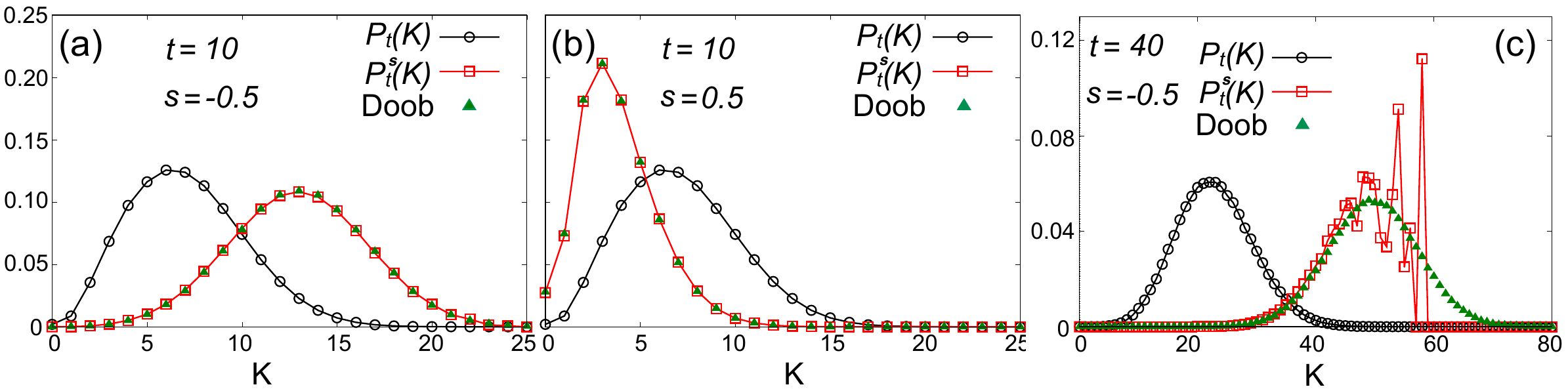}
\caption{
(a) Biasing towards an active dynamics ($s=-0.5$) for $t=10$. Black circles correspond to the numerical probability distribution of the photon-emission outcome for the original dynamics of three-qubit system. Red squares stand for the biased probability $P^s_t(K)$ obtained from the original distribution $P_t(K)$ while green triangles are the numerical results obtained with the time-dependent Doob dynamics. (b) Analogous results when biasing towards an inactive dynamics ($s=0.5$). (c) Same results biasing towards an active dynamics ($s=-0.5$) for $t=40$. The apparent discontinuous behavior of $P_t^s(K)$ stems from the undersampling of the far tails of the original distribution $P_t(K)$ due to the exponential suppression in time of atypical events. Green triangles are the numerical results obtained with the time-dependent Doob dynamics.}
\label{figS1}
\end{figure*}

This procedure, which is equivalent to the \emph{gauge transformation} used for the classical case in Ref.\ \cite{Garrahan2016}, defines a class of time-dependent maps generating the same ensemble as $\mathcal{L}_{\vec{s}_0}$. The Doob generator is then obtained by \emph{fixing the gauge}, i.e. choosing $G_t$, such that $\tilde{\mathcal{L}}_t^*[{\bf 1}]=0$. This leads to $\partial_t \left( G_t^\dagger G_t \right) = - \mathcal{L}_{\vec{s}_0}^*[G_t^\dagger G_t]$, that we solve with final condition $G_{t_{\rm f}}^2={\bf 1}$, obtaining $G_t = 
\sqrt{e^{(t_{\rm f} - t) \mathcal{L}^*_{\vec{s}_0}}[{\bf 1}]}$.
This dynamics can be written in a time-dependent Lindblad form with jump operators $\tilde{L}_\mu = \e^{f_\mu(\vec{s}_0)/2} G_t L_\mu G_t^{-1} $ and the time-dependent Hamiltonian, 
\begin{equation}
\tilde{H} =\frac{1}{2}\left[ G_t \left(H-\frac{i}{2}\sum_{\mu=1}^{N_J}L_\mu^\dagger L_\mu\right) G_t^{-1} + i \partial_t G_t G_t^{-1}+{\rm h.c.}\right]
\end{equation} 
reducing, for long observation times $t_{\rm f} \to \infty$, to the time-independent result \ers{TiltDoob}{HDoob}. In 
the next section we present numerical results demonstrating the validity of our finding on the finite-time Doob dynamics.

\smallskip
\noindent  {\bf \em Examples of the finite-time quantum Doob transform.} In the following we provide some examples in which we use the finite-time quantum Doob transform to generate the probability distribution of the emitted photons when biasing 
the system to a more active or inactive behavior with respect to the typical value.
In all cases we take 
$V=\gamma=\Omega=1$ and we start from the initial configuration where the three qubits are in the up state.

In Fig.~\ref{figS1}(a) we display the same data as Fig.~1, where we bias the original system towards a more active dynamics by taking a negative value of the tilting field ($s=-0.5$). We 
observe how the photon-emission statistics up to $t=10$ obtained via the Doob dynamics (green triangles) is equal to  the biased probability distribution $P^s_t(K)=e^{-s K}P_t(K)/Z_t(s)$ (red squares) obtained from the original probability distribution $P_t(K)$ (black circles). Both the Doob and the original distribution have been obtained by sampling $5\times 10^5$ trajectories generated via quantum jump Monte Carlo for the (time-dependent) finite-time Doob dynamics (with ${\tilde H}$ and ${\tilde L}$) and the original dynamics (with $H$ and $L$) respectively. The Doob dynamics gives rise to an average photon-emission $\la K \ra\sim 13$ which is larger than the average of the original system $\la K \ra\sim 7$. Same results are displayed in Fig.~\ref{figS1}(b) when biasing the system towards a more inactive dynamics by taking a positive tilting or biasing field ($s=0.5$). In this case the average of the modified system is shifted to $\la K \ra \sim 4$.

In Fig.~\ref{figS1}(c) we condition the original system to be more active ($s=-0.5$, as in Fig.~\ref{figS1}(a)), but we study the statistics of the photons emitted up to a larger time $t=40$. Again, the number of trajectories to sample the original and the finite-time Doob dynamics is equal to $5\times 10^5$. However, we observe how in this case the biased probability $P^s_t(K)$ obtained from the sampled probability of original dynamics is not meaningful. This is due to the fact that as time increases the tails of the original distribution (black circles) are not properly sampled, i.e. obtaining quantum rare trajectories from simulation of the original dynamics becomes more and more difficult. This problem is overcome by considering the quantum trajectories via the Doob dynamics (green triangles) which allows us to make typical the rare trajectories of the original system, thus having a faithful sampling even in the far tails of the distribution.


\smallskip

\noindent {\bf \em Conclusions.} 
We have derived by means of dynamical large deviations techniques a constructive way for making rare dynamical fluctuations typical in open quantum systems. Our results here open up the possibility of tailoring open quantum systems in order to obtain a desired statistics of emissions on demand. 

\begin{acknowledgments}
The research leading to these results has received funding from the European Research Council under the European Union's Seventh Framework Programme (FP/2007-2013) / ERC Grant Agreement No. 335266 (ESCQUMA) and the EPSRC Grant No. EP/M014266/1. We are also grateful for access to the University of Nottingham High Performance Computing Facility.
\end{acknowledgments}

\bibliographystyle{apsrev4-1}
\bibliography{3qDoob-biblio}

\onecolumngrid
\newpage

\renewcommand\thesection{S\arabic{section}}
\renewcommand\theequation{S\arabic{equation}}
\renewcommand\thefigure{S\arabic{figure}}
\setcounter{equation}{0}

\begin{center}
{\Large Making rare events typical in Markovian open quantum systems: Supplemental Material}
\end{center}

\section*{Explicit form of $\tilde{\mathcal{W}}_{\vec{s},\vec{s}_0}$}
We show how from the definition of the map $\tilde{\mathcal{W}}_{\vec{s},\vec{s}_0}$ it is possible to obtain the form given in (7). To shorten the notation it is useful to define
$$
W=\ell_{\vec{s}_0}^{1/2}\Big(H-\frac{i}{2}\sum_{\mu=1}^{N_J}L_\mu^\dagger L_\mu\Big)\ell_{\vec{s}_0}^{-1/2}\, .
$$
Since $\theta(\vec{s}_0)$ is the eigenvalue of the dual map $\mathcal{L}_{\vec{s}_0}^*$ with respect to the left eigenmatrix $\ell_{\vec{s}_0}$, we can write $\theta(\vec{s}_0)=\ell_{\vec{s}_0}^{-1/2}\, \mathcal{L}^*_{\vec{s}_0}[\ell_{\vec{s}_0}]\, \ell_{\vec{s}_0}^{-1/2}$, so that  
$$
\theta(\vec{s}_0)X=\frac{1}{2}\Big\{\ell_{\vec{s}_0}^{-1/2}\, \mathcal{L}^*_{\vec{s}_0}[\ell_{\vec{s}_0}]\, \ell_{\vec{s}_0}^{-1/2},X\Big\}\, .
$$
Writing term by term the operator $\ell_{\vec{s}_0}^{-1/2}\, \mathcal{L}^*_{\vec{s}_0}[\ell_{\vec{s}_0}]\, \ell_{\vec{s}_0}^{-1/2}$, we have
\begin{equation}
\begin{split}
\ell_{\vec{s}_0}^{-1/2}\, \mathcal{L}^*_{\vec{s}_0}[\ell_{\vec{s}_0}]\, \ell_{\vec{s}_0}^{-1/2}=iW^\dagger -iW+\sum_{\mu=1}^{N_J}\e^{f_\mu(\vec{s}_0)}\Big(\ell_{\vec{s}_0}^{-1/2}L^\dagger_\mu\ell_{\vec{s}_0}^{1/2}\Big)\Big(\ell_{\vec{s}_0}^{1/2}L_\mu\ell_{\vec{s}_0}^{-1/2}\Big)\, ,
\end{split}
\label{P1}
\end{equation}
and with the definition $\tilde{L}_\mu^{\vec{s}_0}=\e^{\frac{1}{2}f_\mu(\vec{s}_0)}\ell_{\vec{s}_0}^{1/2}L_\mu\ell_{\vec{s}_0}^{-1/2}$, it can be written as
$$
\ell_{\vec{s}_0}^{-1/2}\, \mathcal{L}^*_{\vec{s}_0}[\ell_{\vec{s}_0}]\, \ell_{\vec{s}_0}^{-1/2}=iW^\dagger -iW+\sum_{\mu=1}^{N_J}\tilde{L}_\mu^{\vec{s}_0\dagger}\tilde{L}_\mu^{\vec{s}_0}\, .
$$
On the other hand, since $f_\mu(\vec{s})$ are linear functions of $\vec{s}$ such that $f(\vec{s}=0)=0$, one has $\e^{f_\mu(\vec{s}+\vec{s}_0)}=\e^{f_\mu(\vec{s})}\e^{f_\mu(\vec{s}_0)}$, one can also expand $\ell_{\vec{s}_0}^{1/2}\mathcal{L}_{\vec{s}+\vec{s}_0}[\ell_{\vec{s}_0}^{-1/2}X\ell_{\vec{s}_0}^{-1/2}]\ell_{\vec{s}_0}^{1/2}$ as follows
\begin{equation}
\begin{split}
\ell_{\vec{s}_0}^{1/2}\mathcal{L}_{\vec{s}+\vec{s}_0}[\ell_{\vec{s}_0}^{-1/2}X\ell_{\vec{s}_0}^{-1/2}]\ell_{\vec{s}_0}^{1/2}=-iWX+iXW^\dagger +\sum_{\mu=1}^{N_J}\e^{f_\mu(\vec{s})}\tilde{L}_\mu^{\vec{s}_0}X\tilde{L}_\mu^{\vec{s}_0\dagger }\, .
\end{split}
\label{P2}
\end{equation}
Using relation \eqref{P1},\eqref{P2}, and recalling the definition 
$$
\tilde{\mathcal{W}}_{\vec{s},\vec{s}_0}[X]=\ell_{\vec{s}_0}^{1/2}\mathcal{L}_{\vec{s}+\vec{s}_0}\left[\ell_{\vec{s}_0}^{-1/2}\, X\, \ell_{\vec{s}_0}^{-1/2}\right]\ell_{\vec{s}_0}^{1/2}-\theta(\vec{s}_0)X \, ,
$$
we can write
\begin{equation}
\tilde{\mathcal{W}}_{\vec{s},\vec{s}_0}[X]=-i\left(\frac{W+W^\dagger}{2}\right)X+iX\left(\frac{W+W^\dagger}{2}\right)+\sum_{\mu=1}^{N_J}\left(\e^{f_\mu(\vec{s})}\tilde{L}_\mu^{\vec{s}_0}X\tilde{L}_\mu^{\vec{s}_0\dagger}-\frac{1}{2}\left\{\tilde{L}_\mu^{\vec{s}_0\dagger}\tilde{L}_\mu^{\vec{s}_0},X\right\}\right)\, .
\end{equation}
Noticing that $\tilde{H}_{\vec{s}_0}$ in the main text is given by $\tilde{H}_{\vec{s}_0}=\frac{W+W^\dagger}{2}$, we recover the expression of $\tilde{\mathcal{W}}_{\vec{s},\vec{s}_0}[X]$ in equation (7).

\end{document}